# Structural responses incipient to pressure-driven antiferromagnetic quantum critical point of van der Waals heavy-fermion metal CeSiI

Hanming Ma[1,2=], Tong Shi[1,3=], Wenhao Li[4=], Qingxin Dong[1,2], Xiaoli Ma[1,2], Shaoheng Ruan[1,2], Zhongjin Wu[1,2], Pengtao Yang[1,2], Zhaoming Tian[3], Jianping Sun[1,2], Yoshiya Uwatoko[5, 6], Xiaohui Yu[1,2*], Hechang Lei[4*], Bosen Wang[1,2*], and Jinguang Cheng[1,2*]

[1] Beijing National Laboratory for Condensed Matter Physics and Institute of Physics, Chinese Academy of Sciences, Beijing 100190, China

[2] School of Physical Sciences, University of Chinese Academy of Sciences, Beijing 100049, China

[3] Wuhan National High Magnetic Field Center and School of Physics, Huazhong University of Science and Technology, Wuhan 430074, China

[4] School of Physics and Key Laboratory of Quantum State Construction and Manipulation (Ministry of Education), Renmin University of China, Beijing, 100872, China

[5] Department of Advanced Materials Science, Graduate School of Frontier Sciences, University of Tokyo, Kashiwa, Chiba 277-8581, Japan

[6] Department of Natural Sciences, Faculty of Science and Engineering, Tokyo City University, Setagaya-ku, Tokyo 158-8557, Japan

*Corresponding authors: bswang@iphy.ac.cn (BSW); yuxh@iphy.ac.cn (XHY); hlei@ruc.edu.cn (HCL); jgcheng@iphy.ac.cn (JGC)

CeSiI is a van der Waals heavy-fermion metal recently found to exhibit unconventional superconductivity near a pressure-induced antiferromagnetic quantum critical point (QCP) at $P_c \approx 6$ GPa. Here, we report a comprehensive single-crystal X-ray diffraction study of CeSiI under high pressures up to 8.3 GPa at room temperature, revealing subtle structural responses that precede pressure-driven QCP. We find that the unit-cell volume decreases smoothly upon compression without showing any structural phase transition in the investigated pressure range. Intriguingly, we observe abrupt and concurrent anisotropic responses of the lattice parameters around $P_c \approx 6$ GPa, *i.e.*, the *a*-axis contracts while the *c*-axis enlongated suddenly, with the unit-cell volume smoothily varies with pressure. Structural refinements further show that these lattice anomalies primarily originate from changes of Ce-Ce and Ce-Si bond lengths, as well as a flattening of the inner honeycomb Si layer within the CeSiI monolayer around $P_c$. Our findings establish an interesting case linking pressure-driven electronic transition of QCP at low temperatures to incipient structural responses at room temperature, thereby providing fresh insight into the pressure-temperature phase diagram of CeSiI.

## Introduction

Low-dimensional heavy-fermion materials have attracted increasing attention, primarily because reduced dimensionality enhances magnetic fluctuations and promotes complex anisotropic interactions and competing orders. These effects give rise to intriguing physical phenomena and offer new routes for tuning physical properties [1-4]. In such systems, the lattice is strongly coupled to a variety of cooperative and competing low-energy excitations [5-12], including the coherent Kondo hybridization state, RKKY-mediated long-range magnetic order, charge or spin density waves, and unconventional superconductivity (SC). This intimate coupling underscores the crucial importance of structural information and its evolution under external stimuli for understanding these intriguing quantum states [13-17].

In addition, theoretical calculations of strongly correlated *f*-electron systems are often hindered by the complex local Coulomb interactions of 4*f* electrons, making it difficult to reliably predict subtle structural distortions, charge redistribution, and hybridization characteristics from first-principles calculations alone [18-20]. In this context, the determination of detailed structural features in heavy-fermion materials via single-crystal diffraction techniques constitutes an essential and direct approach to deepen our understanding of their intricate phase diagrams and underlying mechanisms, thereby offering valuable guidance for the design of novel low-dimensional heavy-fermion materials [1, 21].

In the present study, we focus on the newly discovered two-dimensional van der Waals (vdW) heavy-fermion metal CeSiI [22-24], which crystallizes in a trigonal crystal system with space group *P*-3m1 at ambient pressure (AP). Its layered structure consists of a central buckled Si honeycomb layer sandwiched between triangular Ce layers and terminated by triangular I layers, with adjacent CeSiI slabs separated by vdW gaps, as illustrated in Fig. 1(a). At AP, CeSiI exhibits the onset of Kondo coherence below $T^* \approx 50$ K, followed upon further cooling by an antiferromagnetic (AF) transition at $T_N \approx 7.5$ K [22, 23]. The combination of various ambient-pressure properties including a highly frustrated AF order, incommensurate Kondo hybridization [22, 23, 25], the local coordination environment of the Ce ions [21, 26, 27], and pronounced non-Fermi-liquid behavior [23] have established CeSiI as a rare platform for exploring the Kondo-RKKY competition and exotic quantum criticality in the two-dimensional limit [10-12, 22-24]. Our recent high-pressure (HP) investigations reveal that $T^*$ initially decreases with increasing pressure, reaches a minimum near 6 GPa, and then increases upon further compression [24]. In contrast, $T_N$ decreases monotonically and is completely suppressed near $P_c \approx 6$ GPa, where a superconducting state emerges as the AF order collapses. The resulting pressure-temperature phase diagram is summarized in Fig. 1(b).

Despite these rich macroscopic properties, a reliable microscopic picture of the atomic and electronic structure underlying these exotic phenomena in CeSiI remains lacking. In this regard, accurate information on the crystal structural and electron density distribution as a function of pressure and/or temperature is indispensable [21, 26, 27]. As demonstrated in our previous work on $CeNiC_2$, the evolution of local coordination environment of the Ce ions is intimately linked to the Kondo-RKKY competition, confirming that the atomic-scale structural information provides a vital microscopic perspective on such systems [28]. Meanwhile, a particularly important question arises: whether certain structural features can appear at high temperatures incipient to the pressure-driven magnetic QCP and correlate with the low-temperature electronic phase diagram [10-12, 22-24].

Motivated by these considerations, we performed a systematic structural study of CeSiI, using high-precision single-crystal X-ray diffraction (SXRD) measurements up to 8.32 GPa. Our results reveal a subtle yet clear anisotropic response of lattice parameters across $P_c \approx 6$ GPa at room temperature due to an unusual modification of local coordination symmetry of the Ce atoms within individual CeSiI layers, despite of a smooth variation with pressure of unit-cell volume. The present work provides valuable and reliable structural information for achieving a deeper understanding of the temperature-pressure phase diagram of CeSiI.

## Experimental methods

High-quality single crystals of CeSiI were synthesized by high-temperature solid-state reaction method as described in Refs. [22-24]. SXRD measurements were performed at room temperature using a Rigaku XtaLab Synergy R diffractometer equipped with a HyPix detector and Mo-K$\alpha$ radiation ($\lambda = 0.71073$ Å). The crystal structure was solved using Olex2 via direct methods (SHELXT 2018/2) and refined employing full-matrix least squares techniques based on $F^2$ values (SHELXL 2018/3) [29, 30]. High-pressure experiments were carried out using a diamond anvil cell (DAC) with a culet size of 300 $\mu$m. A rhenium metal gasket, initially 250 $\mu$m thick, was pre-indented to a thickness of ~ 70 $\mu$m, and a 200 $\mu$m diameter hole was subsequently drilled into it. A carefully selected CeSiI single crystal, with typical dimensions of ~ $100 \times 100 \times 40$ $\mu m^3$, was loaded into the DAC alongside a piece of ruby used as a pressure manometer. Silicone oil served as the pressure-transmitting medium. The pressure was determined by the ruby fluorescence calibration method [31].

## Experimental results

### Crystal structure and lattice compression of CeSiI

Fig. 2(a) shows the SXRD pattern of CeSiI at AP, which can be successfully refined with the trigonal *P*-3m1 space group, yielding lattice parameters $a$ = 4.1820 (2) Å, $c$ = 11.7149 (7) Å and unit-cell volume $V$ = 177.43 (2) Å$^3$. Theres values are in excellent agreement with previous reports [22, 23]. In addition, the sharp, well-defined diffraction spots confirm the high crystalline quality of the studied sample. We then proceeded to applying pressures and collected SXRD patterns of CeSiI under various pressures up to 8.3 GPa. Some representative patterns are shown in Figs. 2(b)-(d). As can be seen, well-defined diffraction peaks are preserved under pressures up to 8.3 GPa. Moreover, all SXRD patterns can be described by the same *P*-3m1 space group, showing no sign of a structural transition in the investigated pressure range. These high-quality SXRD data thus enabled us to extract accurate crystallographic information of CeSiI single crystals, including the lattice parameters, atomic coordinates, bond lengths and angles under various pressures, which are collected in Tables 1 and 2.

The pressure dependences of $a$, $c$, and $V$ are shown in Figs. 2(e)-(g). With increasing pressure, both $a$ and $c$ decrease monotonically, and then exhibit an abrupt anisotropic responses: the $a$-axis shows a sudden contraction, while the $c$-axis displays a concurrent step-like increase at the same pressure of $P_c \approx 6.6$ GPa, as indicated by the vertical dotted line. Despite the pronounced anomalies in the lattice parameters, $V(P)$ decreases smoothly over the studied pressure range without showing any discontinuity around $P_c$, suggestive of a second-order phase transition. For comparison, the relative compressibility $a/a_0$, $c/c_0$ and $V/V_0$ are plotted as a function of pressure in Fig. 2(h). At 8.32 GPa, $a/a_0$ and $c/c_0$ are reduced to 0.95 and 0.90 GPa, respectively. The high compressibility of the $c$-axis should be ascribed to the weakly coupled vdW gap along this direction as confirmed below. It also reflects the intrinsic anisotropic mechanical response in the layered CeSiI structure. As shown in Fig. 2(g), fitting the $V(P)$ data to the Birch-Murnaghan equation of state as below [32]:

$$P = \frac{3B_0}{2}\left[\left(\frac{V_0}{V}\right)^{\frac{7}{3}} - \left(\frac{V_0}{V}\right)^{\frac{5}{3}}\right]\left\{1 + \frac{3}{4}\left(B_0^{'} - 4\right)\left[\left(\frac{V_0}{V}\right)^{\frac{2}{3}} - 1\right]\right\} \quad (1),$$

where $V$ is the volume at a fixed pressure $P$, $V_0$ is the volume at ambient pressure, and $B_0$' is the first derivative of the bulk modulus. The fitting yields a bulk modulus of $B_0$ = 33.3±1.2 GPa and $B_0$' = 3.3, which is consistent with large compressibility typical of 2D materials [33-35].

**Pressure dependence of inter-layer and intralayer distances in CeSiI**

To further resolve the origin of these structural anomalies, we first examined the pressure evolutions of characteristic inter- and intra-layer distances, as marked in Fig. 3(a): namely, the interlayer I-I distance across the vdW gap ($h_{\text{I-I}}^{\text{Inter}}$), the distances between I-I layers ($h_{\text{I-I}}^{\text{Intra}}$) and Ce-Ce layers ($h_{\text{Ce-Ce}}$) within the CeSiI slabs, the height of the buckled Si layer ($h_{Si\text{-}Si}$) in the middle of CeSiI slabs, and the nearest-neighbor in-plane Ce-Ce distance ($d_{Ce\text{-}Ce}$). The pressure dependences of these distances are summarized in Figs. 3(b)-(f). As can be seen, the inter-layer $h_{I\text{-}I}^{Inter}$ decreases smoothly without showing any discontinuity around $P_c$, Fig. 3(b), and it is reduced by nearly 0.6985 Å from 3.3877 Å at AP to 2.6892 Å at 8.3 GPa. In contrast, both $h_{I\text{-}I}^{Intra}$ and $h_{Ce\text{-}Ce}$ first exhibit a rapid reduction above 4 GPa, reach minima near 6 GPa, and then experience a sudden expansion at ~ 6.6 GPa, above which both distances decrease rapidly upon further compression, as seen in Figs. 3(c) and (d). This leads to a total reduction of $h_{I\text{-}I}^{Intra}$ by ~ 0.2164 Å, which is much smaller than the reduction of $h_{I\text{-}I}^{Inter}$. These results confirm that the large compressibility of the $c$-axis mainly originates from the rapid reduction of the vdW gap, while the observed step-like anomaly of $c$-axis at $P_c \approx 6.6$ GP in Fig. 2(f) is attributed merely to the local structural deformations within the CeSiI monolayer, which will be described in the next section. Accordingly, the anomaly of $a$-axis at $P_c \approx 6.6$ GP results from the sudden reduction of the in-plan $d_{Ce\text{-}Ce}$, Fig. 3(e).

It is interesting to note that the height of the inner Si layer displays nonmonotonic evolutions with pressure, Fig. 3(f): $h_{Si\text{-}Si}$ starts to contract at 3.2 GPa, collapses to nearly zero at ~ 6 GPa, and then increases continuously up to 8.32 GPa. As indicated by the vertical dotted line in Figs. 3(c)-(f), the buckled Si layer becomes nearly flattened at the critical pressure where the intralayer distances experience discontinuities. These results indicate that local coordination environment within the CeSiI monolayer undergoes cooperative modifications upon compression.

**Pressure dependence of local coordination environment of Ce atoms in CeSiI**

As shown in Fig. 4(a), each Ce atom is coordinated by three I and six Si atoms, and the nearest-neighbor Ce and Si atoms on both sides of the Si layers form a slightly distorted hexagonal bipyramid. The flat triangular I atom layer results in three identical nearest-neighbor Ce-I bond lengths ($d_{Ce\text{-}I} \times 3$). In contrast, the buckled honeycomb Si plane at AP leads to two sets of Ce-Si bond lengths in the hexagonal bypyramid, which we denoted as $d_{Ce\text{-}Si1} \times 3$ and $d_{Ce\text{-}Si2} \times 3$ for the shorter and longer sets, respectively. The Ce-Si-I block units at 0 GPa and 6.6 GPa are illustrate in Fig. 4(b). Fig. 4(c) shows the pressure dependences of these bond lengths. As can be seen, $d_{Ce\text{-}I}$ exhibits an almost linear pressure dependence over the studied pressure range without discernable anomaly around $P_c$. In contrast, the difference between $d_{Ce\text{-}Si1}$ and $d_{Ce\text{-}Si2}$ first enlarges from AP up to 3 GPa and then reduced at higher pressures until they becomes equal at about 6 GPa, above which these two bonds split again. However, the average Ce-Si bond length $\langle d_{Ce\text{-}Si} \rangle = (d_{Ce\text{-}Si1} + d_{Ce\text{-}Si2})/2$ exhibits a nearly linear shrinkage with pressure as the $d_{Ce\text{-}I}$ does. Such smooth variations of Ce-I and Ce-Si bond lengths cannot explain the sudden expansion of Ce-Ce inter-layer distance, $h_{Ce\text{-}Ce}$, of the hexagonal bipyramid at $P_c \approx 6.6$ GPa. This leads us to further examine the bond angles surrounding the Ce atoms, i.e., $\theta_{\text{I-Ce-I}}$ and $\theta_{\text{Si-Ce-Si}}$ as defined in Fig. 4(a). As can be seen in Fig. 4(c), these two bond angels undergo sudden reduction at $P_c$, which can rationalize the abrupt expansion of $h_{\text{Ce-Ce}}$, or the $c$-aixs.

These results demonstrate that the six surrounding Si atoms shift to adopt nearly equivalent coordination positions around $P_c$, i.e., $z$(Si) ≈ 0.5 in Table 1, consistent with the flattening of the Si layer shown in Fig. 3(f). This changes the $CeSi_6$ hexagonal pyramid from a slightly distorted one with three short and three long edges to a regular one with six equivalent edges. Meanwhile, the concomitant expansion of $h_{\text{Ce-Ce}}$ and the reduction of $\theta_{\text{I-Ce-I}}$ and $\theta_{\text{Si-Ce-Si}}$ around $P_c$ also explain the sudden jump of $c$-axis. The sudden expansion of $h_{\text{Ce-Ce}}$ along the $c$-axis is compensated by the step-like contraction of in-plane Ce-Ce distance ($d_{\text{Ce-Ce}}$) around $P_c$ as shown in Fig. 3(e), leading to the smooth pressure dependence of unit-cell volume. As mentioned above, instead of structural transition, such anisotropic structural changes should have an electronic origin.

To gain some insight, we tentatively examine the evolution of the electron density near the Ce ions based on the structural data [29, 30]. Figure 5 presents electron-density map sections around the Ce ion site, parallel to the $ab$ plane under different pressures. As illustrated in Fig. 5(d), the electron density ~ 0.5 Å away from the Ce center exhibits a clear threefold symmetry at low pressures, transforms into a sixfold-symmetric distribution near 6.61 GPa, and reverts to threefold symmetry above 8.3 GPa. The pronounced variation of the electron density occurs along the nearest-neighbor Ce–I and Ce-Si directions, suggestive of a possible charge redistribution near 6.6 GPa. It should be noted, however, that these observations are based on room-temperature SXRD and call for further verification through low-temperature experiments in future.

## Discussions

As previously reported [24], the temperature-pressure phase diagram of CeSiI exhibits an anomalous decrease in both $T_N$ and $T^*$ with increasing pressure, a behavior distinct from that of conventional 3D heavy-fermion systems. Our high-pressure SXRD results suggest that anisotropic lattice contraction might play a key role in driving this phenomenon. In the present work, we observed pronounced structural modifications at room temperature near the critical pressure $P_c$ ~ 6.6 GPa, where the antiferromagnetic QCP is accessed at low temperatures. While the overall space-group symmetry remains unchanged across the investigated pressure range, the lattice parameters $a$ and $c$ exhibit contrasting pressure responses within a narrow interval of 5.9-6.6 GPa.

Detailed analyses of the inter-layer spacing, bond lengths and angles reveal that these lattice anomalies originate from local structural distortions within the Ce-Si-I monolayer. Specifically, the interatomic and inter-layer distances exhibit a pronounced anisotropic pressure dependence between the in-plane and out-of-plane directions, leading to a significant alteration of the local coordination environment surrounding the Ce atoms. Most notably, the collapse of the Si-Si layer spacing to nearly zero signifies a transformation of the corrugated Si honeycomb into a flat atomic layer within the *ab* plane. Together, these results indicate that the anomalous structural responses stem directly from modifications to the interatomic coordination environment, which in turn points to a possible modification of the electronic structure.

As these anomalous structural responses take place at similar pressure as the antiferromagnetic QCP and the emergence of unconventional superconductivity in CeSiI, our results underscore the dominant role of anisotropic interatomic interactions around the Ce atoms in governing the pressure evolution of its physical properties. The profound coupling between local structural modifications and quantum criticality observed here echoes phenomena in other prototypical strongly correlated systems. For instance, the emergence of a nematic QCP in iron-based superconductors is inextricably linked to anisotropic lattice distortions [36, 37], while in heavy-fermion compounds like $CeCu_2Si_2$ and $CeRhIn_5$, pressure-induced isostructural volume collapses or anisotropic lattice evolutions play a vital role in modulating *c*-*f* hybridization near the QCP [38-40]. Viewed in this broader context, the observed distinct expansion (contraction) of the nearest-neighbor Ce-Ce distance along the *c*-aixs (*ab*-plane) together with the flattening of the Si layer in CeSiI serves as a vivid example of strong spin-lattice coupling. These results suggest that modifications to the Ce-Si coordination symmetry at room temperature act as an incipient structural factor that drives the electronic transitions at low temperatures, providing new insights to the mechanisms of pressure-driven antiferromagnetic QCP and unconventional superconductivity in low-dimensional heavy-fermion systems.

Previous angle-resolved photoemission spectroscopy (ARPES) measurements by Posey *et al*. revealed that the flat band in CeSiI originates from the localized $4f^1$ electrons and a weak $k_z$ dependence of the Fermi surface [23]. This highly two-dimensional electronic nature suggests that any perturbations in the local in-plane coordination environment can induce profound modifications to the overall Fermi surface. This is basically consistent with the attribution of the main results and phase diagrams to local environmental changes of CeSiI monolayer. Furthermore, theoretical calculations on monolayer CeSiI by Fumega *et al*. demonstrated that structural strain can effectively tune the competition between the nearest-neighbor pseudospin exchange coupling ($J_1$) and the Kondo coupling ($J_K$) [26]. In this context, the sudden shrinkage of Ce-Ce distance within the *ab*-plane observed in our experiments provides compelling evidence for a pressure-induced modification of the $J_K/J_1$ ratio, thereby highlighting a hidden and highly sensitive structure-property relationship in this system. However, it should be noted that our present high-pressure SXRD experiments were carried out at room temperature, where the competitions among various low-energy excited states is quenched. Although one might reasonably expect that the relative changes in lattice parameters upon cooling to the absolute zero are smaller than that induced by the applied pressure, some of the observed phenomena in this study need further investigation. In future work, we plan to conduct in-situ high-pressure SXRD experiments at low temperatures and to perform theoretical calculations to further elucidate the underlying physical mechanisms.

## Summary

We have performed single-crystal X-ray diffraction measurements on CeSiI up to 8.32 GPa. Our results reveal a pronounced anisotropic change of the lattice parameter along the *a* and *c* axes around $P_c \approx 6.6$ GPa. Analyses of interatomic distances further demonstrate highly anisotropic compressibility along the in-plane and out-of-plane directions. Notably, the van der Waals gap compresses smoothly throughout the entire pressure range; while the anomalous pressure dependence of lattice arises entirely from internal deformations within the Ce-Si-I monolayer. A detailed examination of local atomic coordination reveals that the corrugation in the Si honeycomb layer disappeared around $P_c$, leading to concurrent changes in local Ce-Si coordination symmetry. These structural anomalies point to a pronounced modification of electronic structures incipient to the antiferromagnetic QCP of CeSiI at low temperatures. Our study provides a vital microscopic perspective for understanding the pressure-temperature phase diagram of CeSiI.

## Data availability

All data are available from the corresponding author upon reasonable request.

## Acknowledgements

This work was supported by the National Key Research and Development Program of China (Grant Nos. 2024YFA1408400, 2023YFA1607402, 2023YFA1406100, 2021YFA1400200, 2021YFA1401800, 2023YFA1406500, 2022YFA1403800), the National Natural Science Foundation of China (Grant Nos. 12025408, 12074414, 12274459), the CAS President's International Fellowship Initiative (Grant No. 2024PG0003), the Outstanding Member of Youth Promotion Association of CAS (Grant No. Y2022004), and the International Young Scientist Fellowship of the Institute of Physics, CAS (Grant No. 202604). High-pressure single-crystal XRD measurements were performed at the Synergetic Extreme Condition User Facility (SECUF, https://cstr.cn/31123.02.SECUF).

## Competing interests

The authors declare no competing interests.

## Table and Figure Captions

**Table 1**. Unit-cell parameters ($a$, $c$, and $V$), fractional atomic coordinates $z$, equivalent isotropic displacement parameters $U_{eq}$, and the reliability factors $R_1$ after refinement of SXRD of CeSiI[a] at different pressures and room temperature.

**Table 2**. Selected interlayer spacings ($h_{I\text{-}I}^{Inter}$, $h_{I\text{-}I}^{Intra}$, $h_{Ce\text{-}Ce}$, $h_{Si\text{-}Si}$), bond distances ($d_{Ce\text{-}Ce}$, $d_{Ce\text{-}I}$, $d_{Ce\text{-}Si\,1}$, $d_{Ce\text{-}Si2}$), and bond angles ($\theta_{Si\text{-}Ce\text{-}Si}$, $\theta_{I\text{-}Ce\text{-}I}$) of CeSiI at different pressures and room temperature.

**Figure 1** (a, b) Crystal structure of CeSiI. The black dashed lines denote the unit cell. (c) The pressure-temperature phase diagram of CeSiI adopted from [24].

**Figure 2** (a)-(d) Single-crystal X-ray diffraction patterns of CeSiI under selected pressures. Well-defined diffraction peaks are preserved up to 8.32 GPa, demonstrating the retention of good crystallization throughout the experiment. (e)-(h) Pressure dependences of lattice parameters $a$, $c$, and volume $V$, as well as their corresponding compressibilities, respectively. The dash line in (g) shows the *B-M* fit result.

**Figure 3** (a) Schematic illustration of the interlayer distances and Ce-Ce bond lengths of CeSiI under pressure: $h_{I\text{-}I}^{Inter}$ and $h_{I\text{-}I}^{Intra}$ for the interlayer I-I distances across the van der Waals gap and within the CeSiI blocks, $h_{Ce\text{-}Ce}$ and $h_{Si\text{-}Si}$ for the interlayer Ce-Ce and Si-Si distances within the CeSiI blocks, $d_{Ce\text{-}Ce}$ for the nearest-neighbor Ce-Ce bond length in the *ab* plane of CeSiI. (b)-(f) Pressure dependences of the interlayer distances and bond lengths shown in (a). The dash lines are guided to eye.

**Figure 4** (a) Schematic illustration of the Ce-I bond length ($d_{Ce\text{-}I}$), the first- and second-nearest-neighbor Ce-Si distances ($d_{Ce\text{-}Si1}$ and $d_{Ce\text{-}Si2}$), and the bond angles $\theta_{Si\text{-}Ce\text{-}Si}$ and $\theta_{I\text{-}Ce\text{-}I}$ between Ce atom and its first neighbors Si atoms and I atoms within the CeSiI block. (b) The illustration of Ce-Si-I block at 0 GPa and 6.6 GPa. (c) Pressure dependences of $d_{Ce\text{-}I}$, $d_{Ce\text{-}Si\,1}$, and $d_{Ce\text{-}Si2}$. (d) Pressure dependences of $\theta_{Si\text{-}Ce\text{-}Si}$ and $\theta_{I\text{-}Ce\text{-}I}$.

**Figure 5** (a)-(c) The electron density map section near Ce atom of CeSiI parallel to *ab* plane at (a) 0 GPa, (b) 6.61 GPa, and (c) 8.32 GPa, respectively. The color bar beneath qualitatively describe the scale of electron density. (d) The electron density about 0.5 Å away from Ce ion center, in the section illustrated in (a). Starting from the *a* axis, the electron density values were recorded clockwise as a function of the polar angle $\theta$ from 0 to 360 degree. The orange and purple dash lines, both in subpanel (a) and (d), indicate the Ce-Si first nearest neighbor direction and Ce-I direction at 0 GPa.

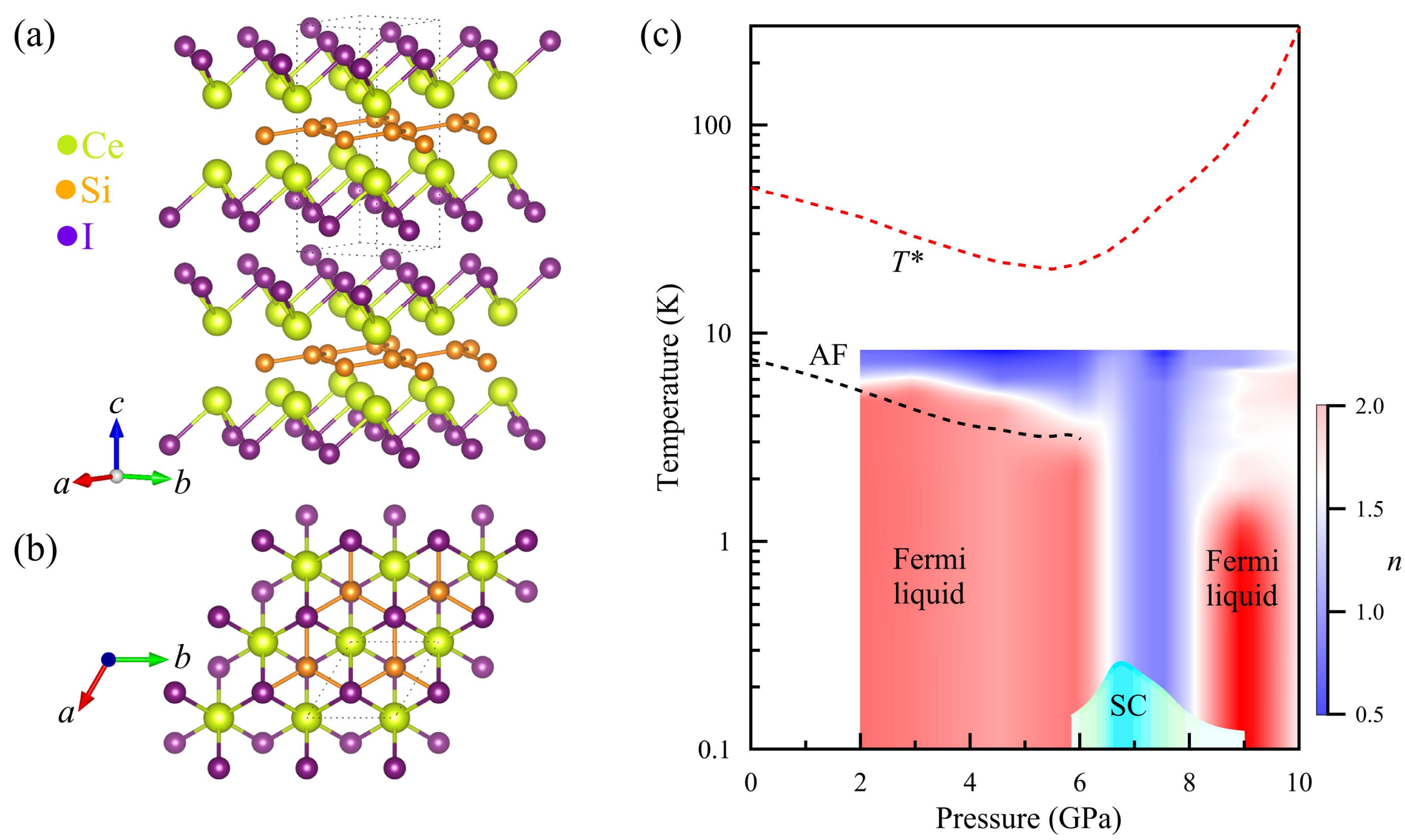
(a)
Ce
Si
I
c
a
b
(b)
b
a
(c)
T*
AF
Fermi liquid
Fermi liquid
SC
Temperature (K)
Pressure (GPa)
n
100
10
1
0.1
0
2
4
6
8
10
2.0
1.5
1.0
0.5

**Figure 1**

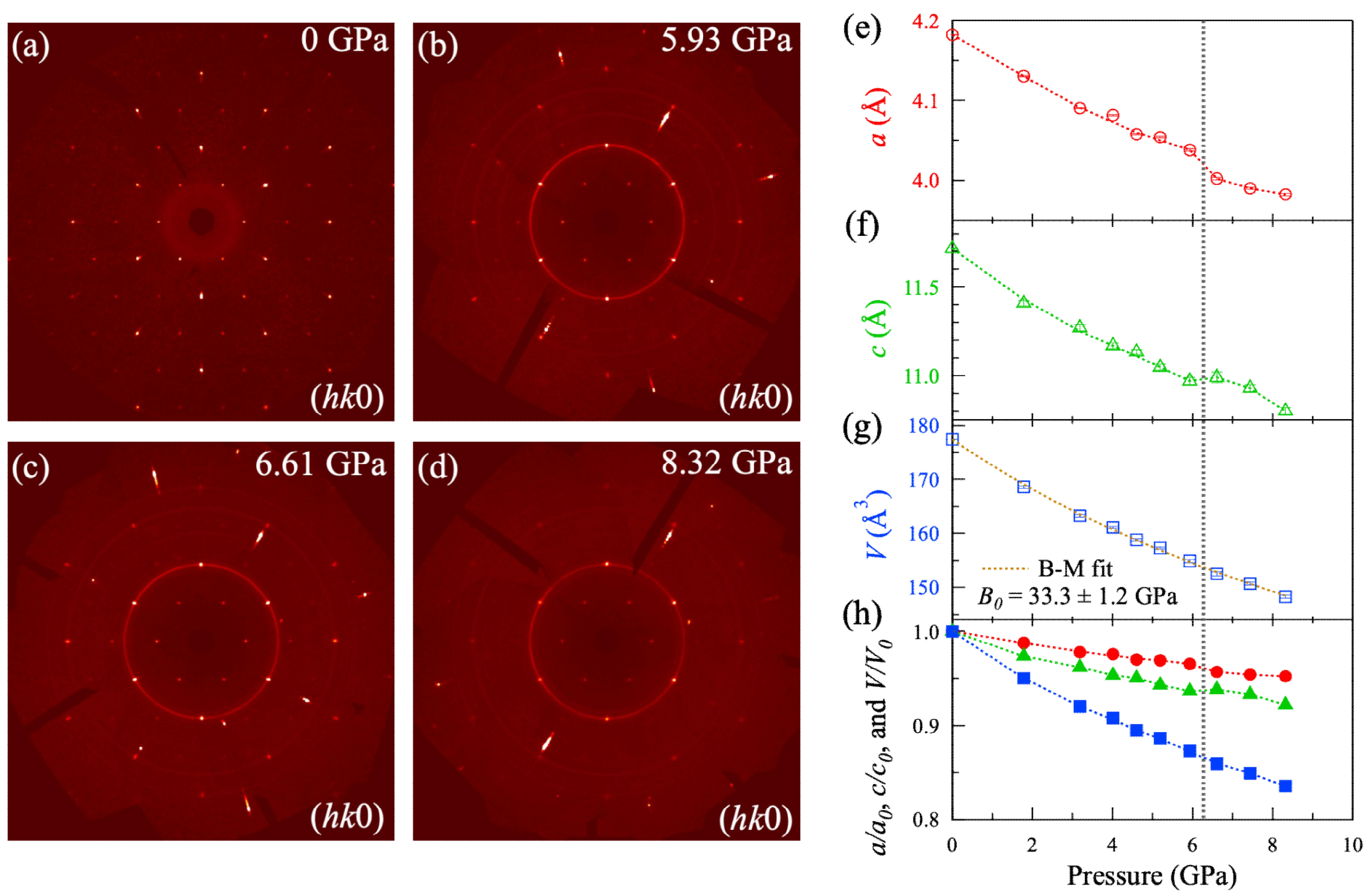

(a)
0 GPa
(hk0)
(b)
5.93 GPa
(hk0)
(c)
6.61 GPa
(hk0)
(d)
8.32 GPa
(hk0)
(e)
a (Å)
(f)
c (Å)
(g)
V (Å3)
B-M fit
B0 = 33.3 ± 1.2 GPa
(h)
a/a0, c/c0, and V/V0
Pressure (GPa)

**Figure 2**

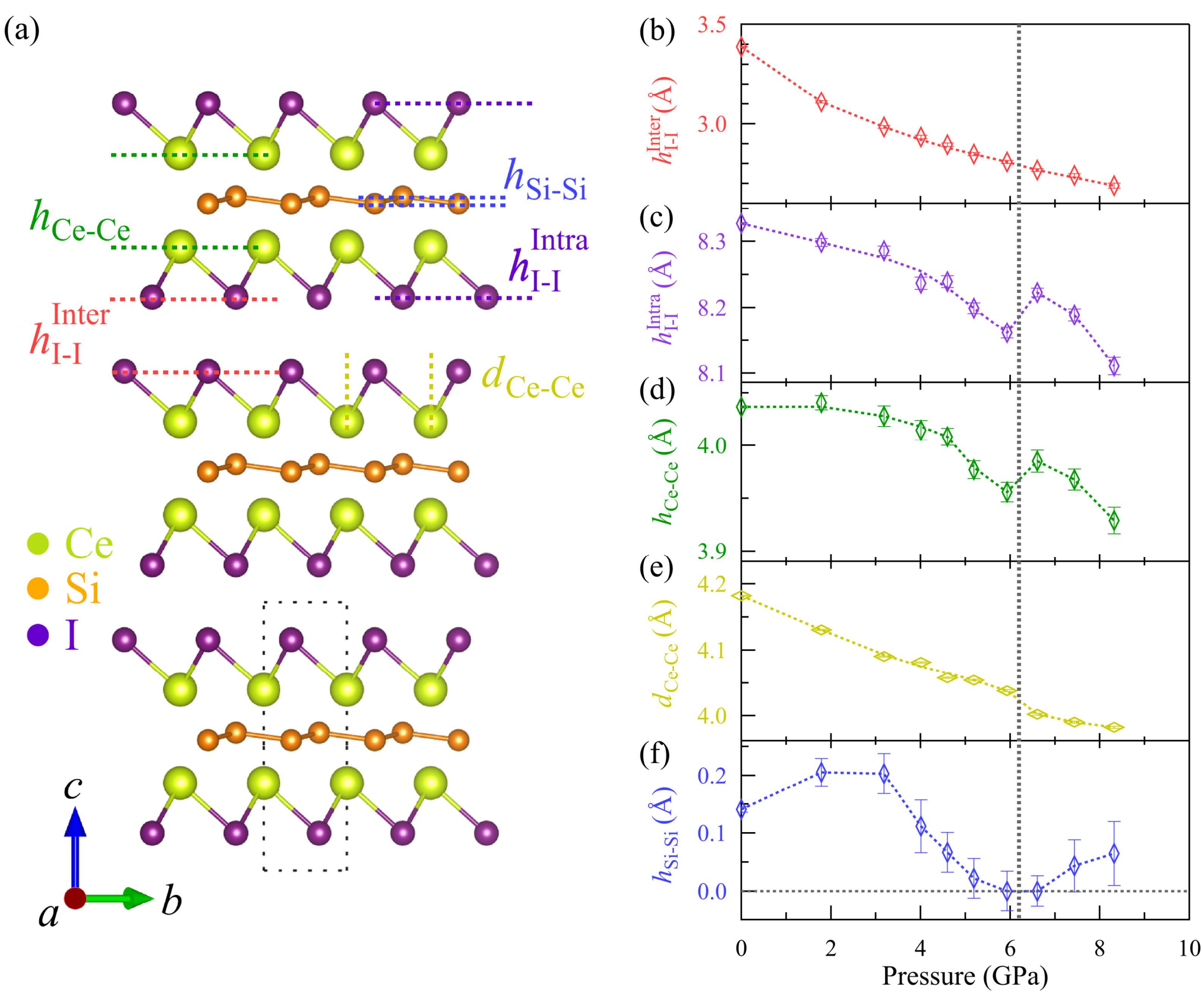

(a)
$h_{\text{Si-Si}}$
$h_{\text{Ce-Ce}}$
$h_{\text{I-I}}^{\text{Intra}}$
$h_{\text{I-I}}^{\text{Inter}}$
$d_{\text{Ce-Ce}}$
Ce
Si
I
c
a
b
(b)
(c)
(d)
(e)
(f)
$h_{\text{I-I}}^{\text{Inter}}$ (Å)
$h_{\text{I-I}}^{\text{Intra}}$ (Å)
$h_{\text{Ce-Ce}}$ (Å)
$d_{\text{Ce-Ce}}$ (Å)
$h_{\text{Si-Si}}$ (Å)
Pressure (GPa)

**Figure 3**

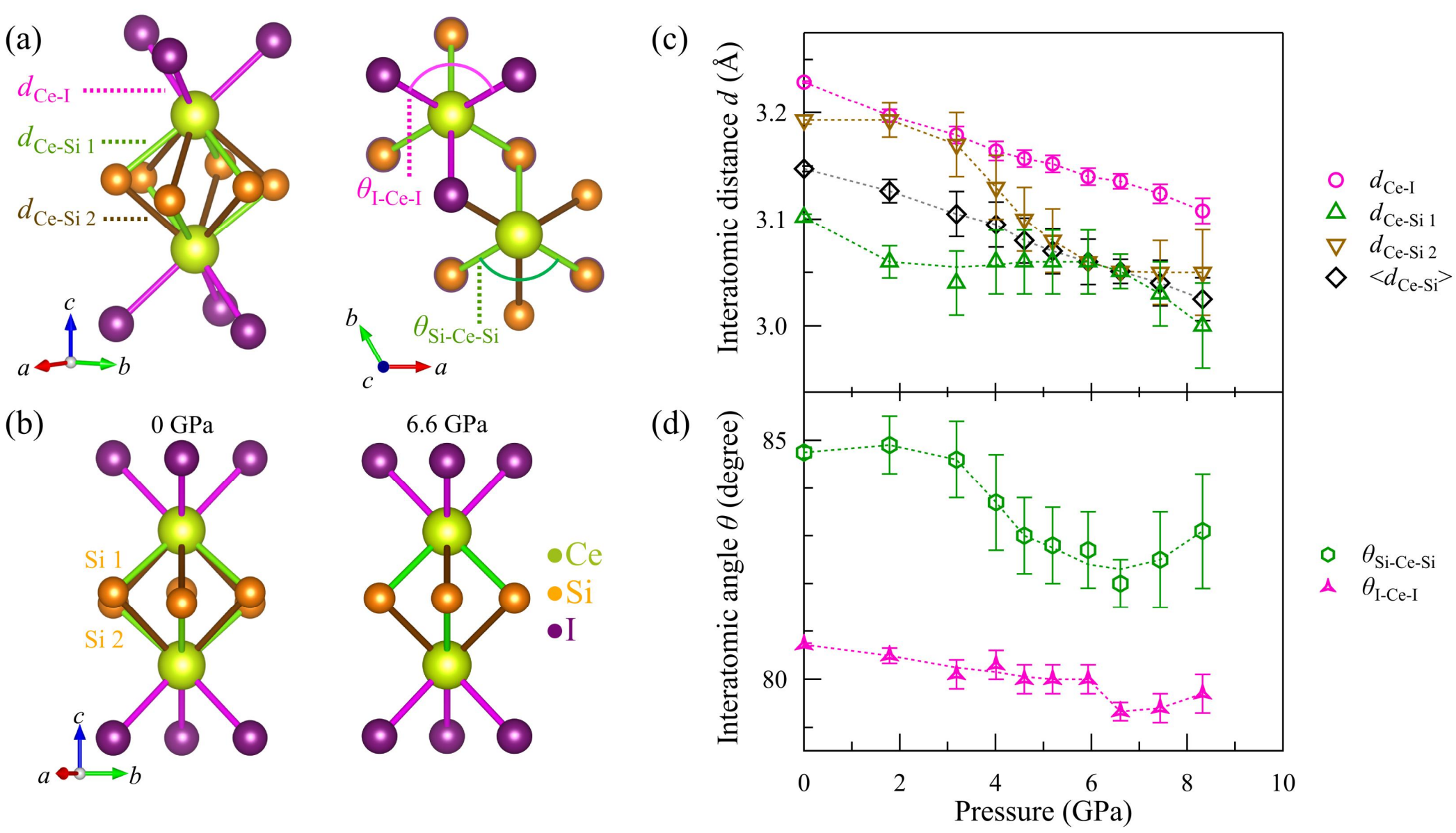

(a)
$d_{Ce-I}$
$d_{Ce-Si\,1}$
$d_{Ce-Si\,2}$
$\theta_{I-Ce-I}$
$\theta_{Si-Ce-Si}$
(b)
0 GPa
6.6 GPa
Si 1
Si 2
Ce
Si
I
(c)
Interatomic distance $d$ (Å)
3.2
3.1
3.0
$d_{Ce-I}$
$d_{Ce-Si\,1}$
$d_{Ce-Si\,2}$
$\langle d_{Ce-Si} \rangle$
(d)
Interatomic angle $\theta$ (degree)
85
80
$\theta_{Si-Ce-Si}$
$\theta_{I-Ce-I}$
0
2
4
6
8
10
Pressure (GPa)

**Figure 4**

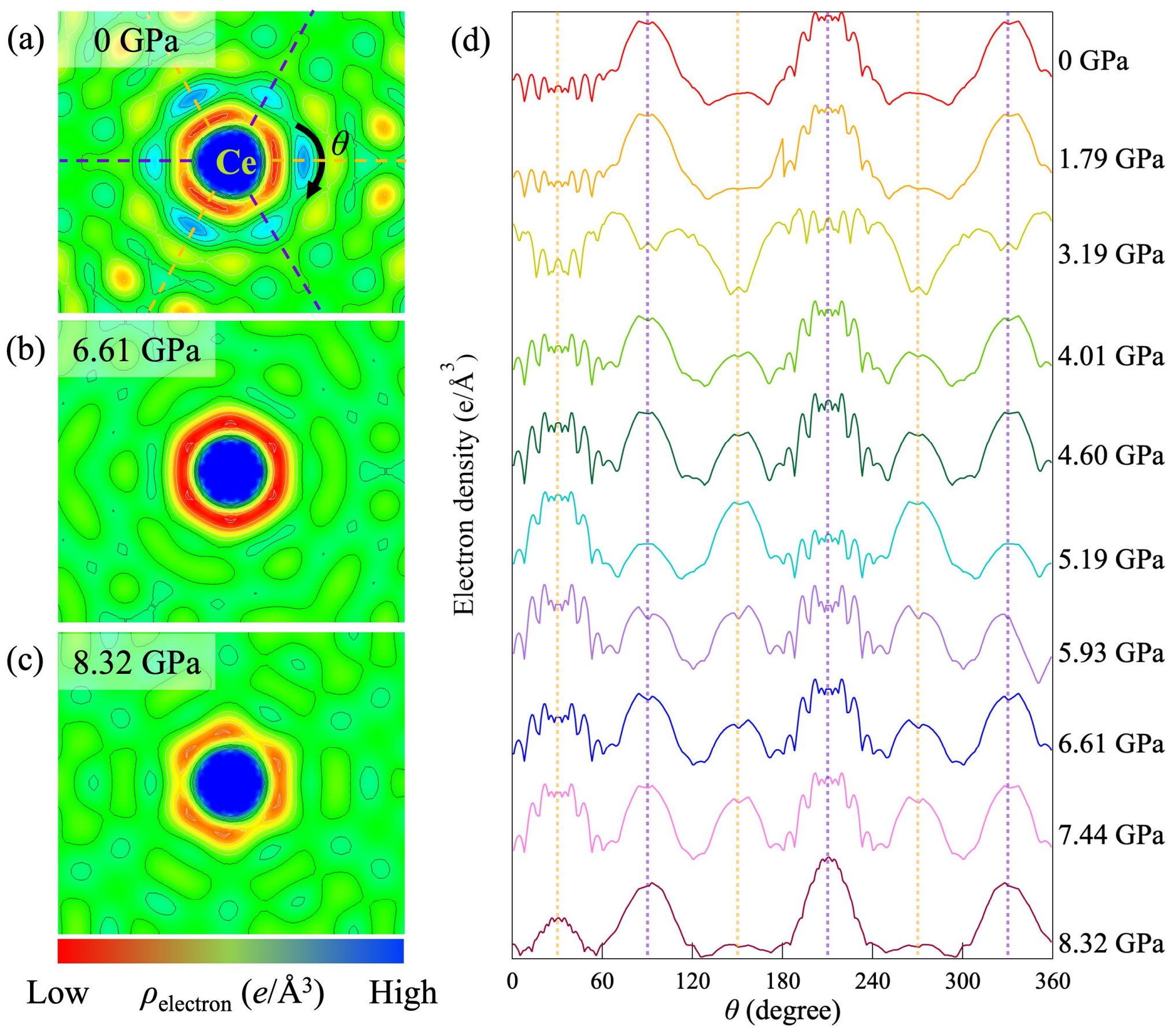
(a)
0 GPa
Ce
θ
(b)
6.61 GPa
(c)
8.32 GPa
Low
ρelectron (e/Å3)
High
(d)
Electron density (e/Å3)
0 GPa
1.79 GPa
3.19 GPa
4.01 GPa
4.60 GPa
5.19 GPa
5.93 GPa
6.61 GPa
7.44 GPa
8.32 GPa
0
60
120
180
240
300
360
θ (degree)

**Figure 5**